\documentstyle[aps,prc,epsfig]{revtex}

\begin{document}

\newcommand{\eee}{{\em et al.}}
\newcommand{\ppp}{\hspace{3em}}
\newcommand{\coes}{\c{c}\~oes }
\newcommand{\cao}{\c{c}\~ao }
\newcommand{\sao}{s\~ao }
\newcommand{\ii }{\'{\i}}
\newcommand{\jps}{$J/ \psi$ }
\newcommand{\nnn}{\noindent}
\newcommand{\beqa}{\begin{eqnarray}}
\newcommand{\eeqa}{\end{eqnarray}}

\title{Meson loop effects on the  pion electromagnetic form factor}
\author{F. Carvalho$^1$, F. O. Dur\~aes$^{1,2}$, F. S. Navarra$^1$ and  M. Nielsen$^1$ }
\address{$^1$Instituto de F\'{\i}sica, Universidade de S\~{a}o Paulo \\ 
C.P. 66318, 05315-970 S\~{a}o Paulo, SP, Brazil\\
$^2$Dep. de F\ii sica, Faculdade de Ci\^encias Biol\'ogicas, Exatas e Experimentais,\\ Universidade 
Presbiteriana Mackenzie, C.P. 01302-907, S\~ao Paulo, SP, Brazil}
\maketitle

\begin{abstract}
In this work we  calculate the meson loop effects on the pion 
electromagnetic form factor and
 perform a comparison with  experimental data. We show that,
 even
 though the meson cloud is not 
the only nonperturbative process to be considered, its 
contribution 
is significant.
\end{abstract}

\section{Introduction}

The pion electromagnetic form factor  has been under intense 
investigation 
both experimentally and theoretically. Yet there are still some
 open questions.  
Perhaps the most intriguing one is: why experimental data in the
 region of 
$Q^2 > 2\, GeV^2$  are still so far away from the perturbative 
QCD 
(pQCD) expectation?  
Or equivalently, what is the nonpertubative dynamics responsible
 for
 this behavior?  
Since long ago it has been suspected that the transition to the 
pQCD regime might
happen at intermediate values of $Q^2$ ($5 < Q^2 < 15\, GeV^2$) 
\cite{bro,kroll,kiss2}. 
However, the data from  Jefferson Laboratory (JLAB) \cite{jefflab} 
are not only far from the pQCD prediction but follow a different 
trend. 

Lattice calculations, in spite of their rapid progress \cite{latt},
 are not yet 
in position to answer the questions above. In order to elucidate
 the subject 
many models have been advanced. A variety of constituent quark 
models has  been 
considered \cite{card,maris,tobias,kiss}. Most of them are based
 on solutions
of the light front Bethe-Salpeter equation  with some 
approximations
. In general, 
they achieve good agreement with data. Along a different line,
  the instanton
model developed in Ref. \cite{fac} was able to give a reasonable 
description of data
in the region $Q^2 > 1\, GeV^2 $. In a similar approach the same
 conclusion was found 
in \cite{mafor}. 

In this work we consider the contribution of the meson cloud 
around 
the pion 
to its eletromagnetic form factor. This  nonperturbative,
 hadronic 
effect was shown to play an  important role in the  
electromagnetic
 form  factor 
of the nucleon \cite{fried,miller,kor,thomas1,mosel}. Recently 
the
 effect of the
pion cloud was considered in the study of the pion form factor in 
the timelike 
region \cite{nach}. The $\rho$ meson, provenient from the photon 
was 
converted into a 
virtual  pion loop. Using the same approach and the same kind of 
vertices, we compute
mesonic loop contribution to the pion form factor in the spacelike 
region.  

The idea of describing a pion with cloud states has already been 
used in \cite{speth} and also 
by us previously 
in \cite{nossoprl}, where we have studied  $D^-/D^+$ asymmetry 
produced
 in 
$\pi p \rightarrow DX$  collisions. In that work we could see 
that the 
inclusion 
of the meson cloud leads  to  a good agreement with experimental
 data. 

In the next section we will review a few formulas used in the
 form 
factor calculations and 
define the dominant  processes. In section III  we discuss the 
inputs for the 
calculations and the numerical results. In section IV we present 
some concluding remarks. 

\section{Loops and the pion form factor}

The pion electromagnetic form factor is defined by:

\beqa
<\pi^-(p')|J_{\mu}|\pi^-(p)>=e(p+p')_{\mu}F_{\pi}(q^2),
\,\,\,\,\,\,\,\,q=p'-p,
\eeqa

\nnn where $J_{\mu}$ is the electromagnetic current. 
As in the case of the nucleon, we can represent the pion as a 
``bare pion'' $|\pi^{-}>_0$  
which can fluctuate into virtual Fock states composed by  a 
pseudo-scalar (P) meson and a 
vector (V) meson, such as:
\begin{eqnarray}
|\pi^->&=&|\pi^->_0+|\pi^- \rho^0>+|\pi^0 \rho^->+|K^- K^{0*}>
+|K^0 K^{-*}>...
\label{fockpion}
\end{eqnarray}

\nnn Thus, we can understand the photon-pion interaction as being 
an interaction between the 
photon and the  constituents of the cloud. Usually the meson 
loop contributions are supposed to be
 dominant at low values of $Q^2$, where the incoming photon does 
not yet resolve quarks inside the
 pion. At higher values of $Q^2$, quark pointlike structures 
should become visible and a
 description of the problem in terms of hadronic degrees of 
freedom should loose validity. 

However, due to the present difficulty in understanding the 
intermediate $Q^2$ data, we consider 
the meson cloud contributions up to $Q^2=6\,GeV^2$.

In Fig. 1  we show the first terms generated  by the expansion 
(\ref{fockpion}): first the interaction between the photon and the 
bare pion (1o) and then   the interaction between the 
photon and states composed by a  pseudoscalar meson and a 
vector meson $|PV>$ (1a-d). The full and open circles represent the 
dressed form factor and the form factor of the bare pion, respectively.

Before presenting the calculations some remarks 
are in order. Quite  generally, the structure, or size, of any hadron depends 
on its quark content, on its meson cloud and on its off-shellness. We want to 
isolate and compute the strength of the meson cloud contribution.  
Looking at Fig. 1 and comparing the pion in the l.h.s. of the 
equality with the pion in a)   we realize that the 
first one is on-shell, whereas the second one is off-shell. Therefore they 
are not exactly the same object and do not have the same structure. Moreover 
the pion in Fig. 1a) 
is  a bare pion  whereas the pion on the l.h.s. of the equality
 is composed by a bare pion plus the sum over all cloud corrections. 
With these considerations 
we want to emphasize that the l.h.s.  pion and the pion within the loop 
are not the same.

\begin{figure}[h]
\begin{center}
\epsfig{figure=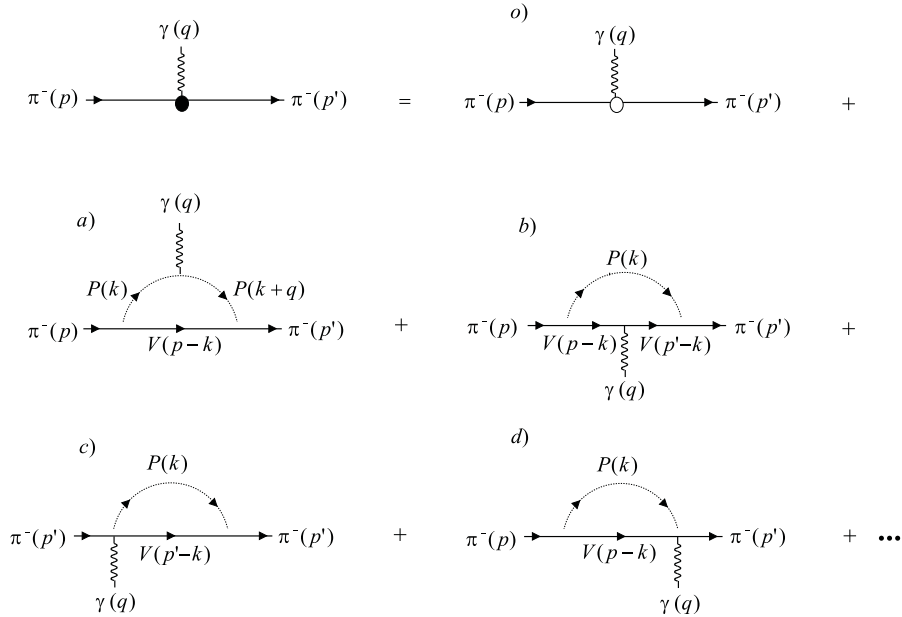,width=12cm} 
\centerline{\caption{Photon-pion interaction in the meson cloud 
approach.}}
\end{center}
\end{figure}

In order to calculate de diagrams of Fig. 1 we start 
from the interaction Lagrangian given by:
\beqa
{\cal L}_{\pi P V}=i\,g_{\pi P V}\biggr(P\,\vec{\tau}
\,V^{\mu}-V^{\mu}\vec{\tau}\,P\biggr)\cdot 
\partial_{\mu}\vec{\pi}
\label{lagr}
\eeqa

\nnn where P and V denote a pseudoscalar and a vector meson 
respectively. From ${\cal L}$ we 
derive the Feynman rules with which we write the vertex function
 associated with diagram (a) of 
Fig. 1: 
\beqa
\Gamma_{\mu}^{(a)}(p,p')=ig^2Q_{P}\int \frac{d^4k}{(2\pi)^4}
\,F(k)\,F(k+q)
\,\Delta(k)\,\Delta(k+q)\,p'_{\beta}\,D^{\alpha\beta}(p-k) 
\,p_{\alpha}\,(2k+q)_{\mu}
\label{diaga}
\eeqa

and with  diagram (b):
\begin{eqnarray}
&&\Gamma_{\mu}^{(b)}(p,p')=ig^2Q_{V}\int \frac{d^4k}{(2\pi)^4}
\,F^2(k)\,\Delta(k)\,p'_{\beta}\,
D^{\theta \beta}(p'-k)\times \nonumber\\
&&\biggr [(p+p'-2k)_{\mu}\,g_{\theta \eta}-(p'-k)_{\eta}
\,g_{\theta \mu}-(p-k)_{\theta}\,g_{\eta 
\mu}\biggr ]
D^{\alpha \eta}(p-k) \,p_{\alpha}
\label{diagb}
\end{eqnarray}

\nnn where  $Q_P$ and $Q_V$ are the charges of the pseudoscalar 
and vector mesons, and $\Delta(k)$
 and $D^{\alpha \beta}(p-k)$ are their propagators.  They are 
given by:
\begin{center}
$D^{\alpha \beta}(p-k)=\left[\displaystyle 
\frac{g^{\alpha \beta}-(p-k)^{\alpha}(p-k)^{\beta}
/m^2_{V}}{(p-k)^2-m^2_{V}}\right]$ \,\,\,\, and
\,\,\,\, $\Delta(k)=\displaystyle 
\frac{1}{k^2-m_P^2}$
\end{center}

\nnn where  $m_P$  and  $m_V$ are the masses of the pseudoscalar 
and vector mesons respectively.

 In order to account in some way for the finite extent of the 
mesons appearing in the loops of
 Fig. 1, we have included form factors in the hadronic vertices.
 For simplicity we have chosen a
 monopole form:
\beqa
F(k)=\frac{m^2_P-\Lambda^2}{k^2-\Lambda^2}
\label{fforma}
\eeqa

\nnn where $\Lambda$ is the cut-off parameter. Although there is
 no rigorous justification for
 this choice, form factors of this type have been used for 
example
 in Ref. \cite{nach}. These 
form factors render all the following loop integrals finite.

\vspace{0.5cm}

In the presence of the electromagnetic field the nonlocal mesonic
 interaction of
 Eqs. (\ref{lagr}) and (\ref{fforma}) gives rise to vertex 
currents.
 In order to maintain gauge 
invariance we introduce the photon field by minimal substitution 
of 
the momentum variable in the 
form factors. This procedure generates  nonlocal seagull vertices
 $(s)$ \cite{ohta}. They are:
\beqa
s^c_{\alpha\mu}=g\,Q_P\,p'_{\alpha}\,(q-2k)_{\mu}
\frac{F(k)}{(k-q)^2-\Lambda^2} 
\label{seagc}
\eeqa

\nnn for diagram (c) and
\beqa
s^d_{\alpha\mu}=g\,Q_P\,p_{\alpha}\,(q+2k)_{\mu}
\frac{F(k)}{(k+q)^2-\Lambda^2} 
\label{seagd}
\eeqa

\nnn for diagram (d). Using the expressions above to write the
 $\Gamma_{\mu}(p,p')$ functions
for diagrams (c) and (d) we find respectively:
\begin{eqnarray}
\Gamma_{\mu}^{(c)_s}(p,p')=i g^2Q_{P}\int \frac{d^4k}{(2\pi)^4}\,
\Delta(k)\,
F^2(k)\,
\frac{1}{(k-q)^2-\Lambda^2}\,
p'_{\beta}\,
D^{\alpha \beta}(p'-k)
\,p'_{\alpha}
\,(q-2k)_{\mu}
\label{diagcs}
\eeqa
and 
\begin{eqnarray}
&&\Gamma_{\mu}^{(d)_s}(p,p')=-i g^2Q_{P}\int 
\frac{d^4k}{(2\pi)^4}
\Delta(k)\,
F^2(k)\,
\frac{1}{(k+q)^2-\Lambda^2}\,
p_{\beta}\,
D^{\alpha \beta}(p-k)\,
p_{\alpha}\,
(q+2k)_{\mu}
\label{diagds}
\eeqa

Due to the presence of the derivative in Eq. (\ref{lagr}), the
 minimal substitution also 
generates  current-vertex couplings $(v)$ even if the mesonic 
form
 factors are not present. They 
are: 
\beqa
v^c=g\,\biggr(Q_D\,F(k-q)+Q_V\,F(k)\biggr)
\label{vertc}\eeqa
for diagram (c) and
\beqa
v^d=-g\,\biggr(Q_D\,F(k+q)+Q_V\,F(k)\biggr)
\label{vertd}\eeqa

\nnn for diagram (d). Repeating the procedure used above, we 
write 
the $\Gamma_{\mu}(p,p')$ 
functions for diagrams (c) and (d) using (\ref{vertc}) and 
(\ref{vertd}):
\begin{eqnarray}
&&\Gamma_{\mu}^{(c)_v}(p,p')= ig^2\int \frac{d^4k}{(2\pi)^4}\,
\Delta(k)\,\,
F(k)\,
p'_{\beta}\,
D^{\beta \mu}(p'-k)
\Biggr \{ Q_{P}\,F(k-q)\,
+ Q_{V}\,F(k)\,\Biggr \}
\label{diagcv}
\eeqa

\nnn and 
\begin{eqnarray}
&&\Gamma_{\mu}^{(d)_v}(p,p')= ig^2\int \frac{d^4k}{(2\pi)^4}\,
\Delta(k)\,\,
F(k)\,
p_{\beta}\,
D^{\beta \mu}(p-k)
\Biggr \{ Q_{P}\,F(k+q)\,
+ Q_{V}\,F(k)\,\Biggr \}
\label{diagdv}
\eeqa

In order to obtain the total vertex function $\Gamma_{\mu}$, we
  add (\ref{diaga}), (\ref{diagb}),
 (\ref{diagcs}), (\ref{diagds}),  (\ref{diagcv}) and
 (\ref{diagdv})
  to find:
\beqa
\Gamma_{\mu}(p,p')=F_1(Q^2)p_{\mu}+F_2(Q^2)p'_{\mu}\nonumber
\eeqa

The  pion electromagnetic form factor is associated with the
  term:
\begin{equation}
F^+_{\pi}(Q^2)=\frac{F_1(Q^2)+F_2(Q^2)}{2}
\label{fformapi}
\end{equation}

\vspace{0.5cm}
Since in the case  of the pion it has not been discussed before 
we shall, in the remainder of this
 section, briefly show that the inclusion of the terms 
(\ref{diagcs}), (\ref{diagds}), 
(\ref{diagcv}) and (\ref{diagdv}) is crucial for the
 fullfilment 
of the Ward-Takahashi (W-T) 
identity. The W-T identity reads:
\beqa
q^{\mu}\Gamma_{\mu}=Q_{\pi}\biggr(\Sigma(p)-\Sigma(p')\biggr)
\nonumber
\eeqa

\nnn where
\beqa
\Sigma(p)&=&-ig^2\int \frac{d^4k}{(2\pi)^4}\,\,\,\Delta(k)
\,F^2(k)\,p_{\beta}
\,D^{\alpha \beta}(p-k)\,p_{\alpha}
\eeqa

For  diagram (a) we have:
\beqa
&&q^{\mu}\Gamma_{\mu}^{(a)}=-ig^2\,Q_{P}\int 
\frac{d^4k}{(2\pi)^4}\,\,\,\biggr[\Delta(k)-
\Delta(k+q)\biggr]F(k)\,F(k+q)\,p'_{\beta}
\,D^{\alpha \beta}(p-k)\,p_{\alpha}
\label{wa}
\eeqa

and for  diagram (b):
\beqa
&&q^{\mu}\Gamma_{\mu}^{(b)}=-ig^2\,Q_{V}\int 
\frac{d^4k}{(2\pi)^4}\,\,\,\Delta(k)
\,F^2(k)\,p'_{\beta}\biggr[D^{\alpha\beta}(p'-k)-
D^{\alpha\beta}(p-k)\biggr]\,p_{\alpha}
\label{wb}
\eeqa

For the seagull coupling of  diagrams (c) and (d):
\beqa
&&q^{\mu}\Gamma_{\mu}^{(s)}=-ig^2\,Q_{P}\int 
\frac{d^4k}{(2\pi)^4}\,\,\,\Delta(k)\,F(k) 
\times\nonumber\\
&& \biggr[-p'_{\beta}\,D^{\alpha\beta}(p'-k)\,p'_{\alpha}
\,\biggr(F(k)-F(k-q)\biggr)
+p_{\beta}\,D^{\alpha\beta}(p-k)\,p_{\alpha}\,
\biggr(F(k)-F(k+q)\biggr)\biggr]
\label{ws}
\eeqa

and for the vertex coupling of  diagrams (c) and (d):
\beqa
&&q^{\mu}\Gamma_{\mu}^{(v)}=-ig^2\,Q_{P}\int 
\frac{d^4k}{(2\pi)^4}\,\,\,\Delta(k)\,F(k)\biggr[-F(k-q)
\,p'_{\beta}
\,D^{\alpha\beta}(p'-k)\,q_{\alpha}-F(k+q)\,p_{\alpha}\,
D^{\alpha\beta}(p-k)
\,q_{\beta}\biggr]\nonumber\\
&&-ig^2\,Q_{V}\int \frac{d^4k}{(2\pi)^4}\,\,\,
\Delta(k)\,F^2(k)\biggr[-p'_{\beta}\,D^{\alpha\beta}(p'-k)
\,q_{\alpha}-p_{\alpha}\,D^{\alpha\beta}(p-k)\,q_{\beta}\biggr]
\label{wv}
\eeqa

Adding the contributions (\ref{wa}), (\ref{wb}), (\ref{wv}) and
 (\ref{ws}), it can easily be seen that:
\beqa
q^{\mu}\Gamma_{\mu}&=&-ig^2\,Q_{P}\int \frac{d^4k}{(2\pi)^4}\,
\Delta(k)\,F^2(k)\biggr[p_{\beta}D^{\alpha\beta}(p-k)
\,p_{\alpha}-
p'_{\beta}D^{\alpha\beta}(p'-k)\,p'_{\alpha}\biggr]\nonumber\\
&&-ig^2\,Q_{V}\int \frac{d^4k}{(2\pi)^4}\,\Delta(k)\,F^2(k)
\biggr[p_{\beta}D^{\alpha\beta}(p-k)\,p_{\alpha}-p'_{\beta}
D^{\alpha\beta}(p'-k)\,p'_{\alpha}\biggr]\nonumber\\
&&=Q_{P}\biggr(\Sigma(p)-\Sigma(p')\biggr)+Q_{V}
\biggr(\Sigma(p)-\Sigma(p')\biggr)
\eeqa

Due to charge conservation, $Q_{P}+Q_{V}=Q_{\pi}$, and thus we
 have:
\beqa
q^{\mu}\Gamma_{\mu}=Q_{\pi}\biggr(\Sigma(p)-\Sigma(p')\biggr)
\nonumber
\eeqa

As it can be seen from the last factor in expression (\ref{diaga}), we 
are using a 
pointlike coupling  in the virtual pion - photon - virtual pion (Fig. 1a). 
Since this pion has already some structure, we might introduce a form 
factor in this coupling, given by its quark structure (already present in 
the bare pion). However, we are interested only in the meson cloud 
and not in modelling the bare pion.  The most simple and consistent 
procedure is to take, as we did, a pointlike coupling and keep in mind that 
this will maximize the contribution of the cloud, since any
possible form factor would be a decreasing function of $Q^2$, with maximum 
value equal to one (at $Q^2=0$). In this way we are computing the upper value 
for the cloud contribution. 

Being restricted to the cloud contribution and making no attempt 
to accurately describe the data, we have only one parameter (the cut-off 
Lambda in our meson-meson-meson form factors), which is constrained by 
other types of phenomenological analyses. We have thus little freedom and 
therefore some  predictive power.

\section{Numerical results and discussion}

In order to compute the form factor (\ref{fformapi}), it is 
necessary
 to know the value of the relevant masses, coupling constants
 $(g)$ 
and  cut-off parameters $(\Lambda)$. The $g_{\pi \pi \rho}$, 
 $g_{\pi KK^*}$ coupling constants have been measured 
experimentally
 \cite{zhu}. As the masses are all known, only the cut-off values 
remain as  free parameters. So far there is no  systematic study  
which indicates the most  appropriate cut-off value for the 
vertices
 involving pion-meson-meson fluctuations. Since we are using a 
 monopole parametrization for these vertices, we have chosen  
cut-off
 values close to the corresponding vector meson masses.
In Table I we present all the values used in these calculations.
\begin{figure}[h]
\begin{center}
\centerline
{\epsfig{figure=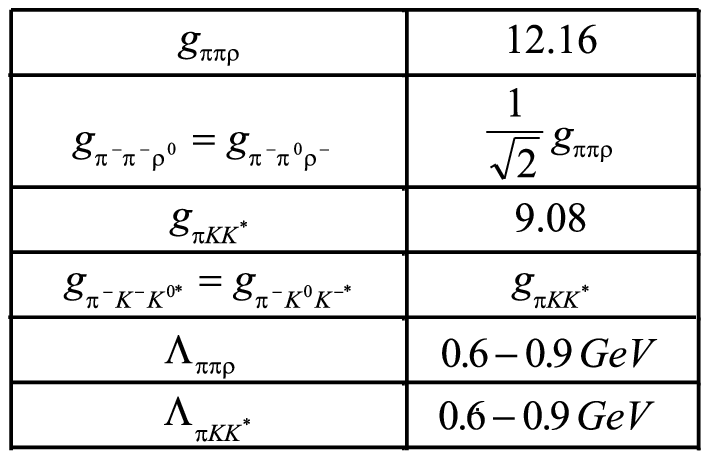,width=7cm}}
\end{center}
\label{table}
\end{figure}
\vspace{-0.7cm}

\begin{center}
{\bf Table I:} \small{Values of coupling constants and cut-off
  used in the vertex functions.}
\end{center}

In order to explicitly show the relative strength of each cloud 
state, we show in Fig. 2 an example of result obtained with (\ref{fformapi}) 
and   a  cut-off fixed to $\Lambda=0.9 \, GeV$ GeV for
the sum of $|\pi \rho>$ states (dashed line), the sum of 
$|KK^*>$ states (dot-dashed line),  and the sum of all of them 
(solid line). We include the experimental data from  the 
$F_{\pi}$ 
collaboration \cite{jefflab}, from Amendolia {\sl et al.}
 \cite{amendolia}, from DESY \cite{desy}  and from Bebeck 
{\sl et al.} \cite{bebeck2}. As it can be seen, all the curves 
rise
, reach a maximum and then drop continuously. The heigth and 
relative strength of each contribution is controlled by the
 masses,
 couplings and cut-off values.
\begin{figure}[h]
\begin{center}
\centerline{\epsfig{figure=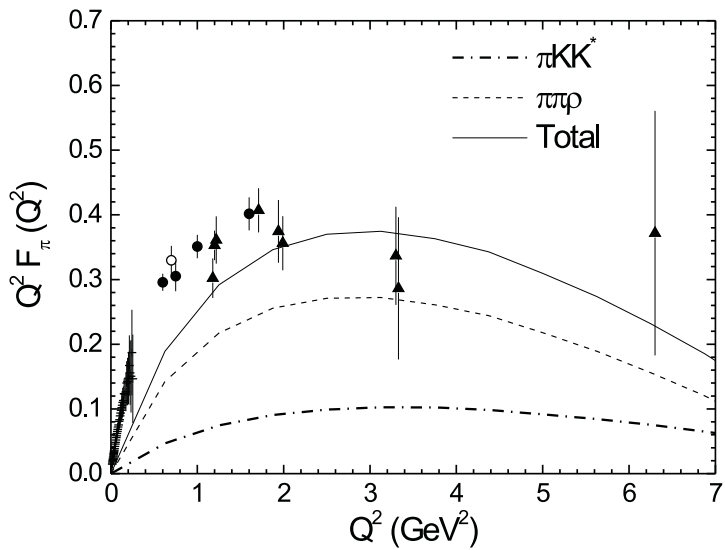,width=11cm}} 
\centerline{\caption{Contribution of the meson cloud states 
to the pion form factor. The black circles, crosses, empty 
circles and  black triangles are the 
experimental data from Refs. \protect\cite{jefflab}, 
\protect\cite{amendolia}, \protect\cite{desy} and 
\protect\cite{bebeck2} respectively. In this figure we 
used $\Lambda_{\pi\pi\rho}$=$\Lambda_{\pi KK^*}$=$0.9 \,GeV$.}}
\label{cut09}
\end{center}
\end{figure}

\begin{figure}[h]
\begin{center}
\centerline{\epsfig{figure=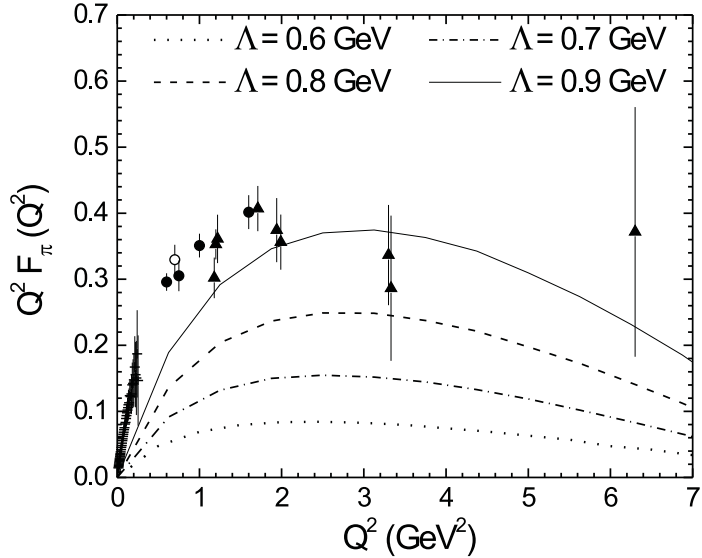,width=11cm}} 
\centerline{\caption{Contribution of the sum of all meson 
cloud states to the quantity $Q^2\,F_{\pi}(Q^2)$, with different cut-off
 choices. In this figure $\Lambda=\Lambda_{\pi\pi\rho}=
\Lambda_{\pi KK^*}$.}}
\end{center}
\label{cutsdif}
\end{figure}

In Fig. 3 we show the contribution of the sum of all the meson cloud 
states to the pion form factor for several cut-off choices. We can 
observe a strong cut-off dependence. Nevertheless it is possible to 
conclude that the cloud contribution is significant specially in 
the region $1\leq Q^2 \leq 6\, GeV^2$. Compared to the case of the 
nucleon electromagnetic form factor 
\cite{miller} the meson cloud plays here a less important role.

For each choice of cut-off parameter in Fig. 3 we have calculated 
the corresponding mean square radius $<r_{\pi}^2>$, according to the 
usual definition:
\begin{equation}
<r_{\pi}^2>  = - 6\,\frac{d F_{\pi}(Q^2=0)}{d Q^2}
\label{raiofinal}
\end{equation}

The results  are presented in Table II.
\begin{figure}[h]
\begin{center}
\centerline
{\epsfig{figure=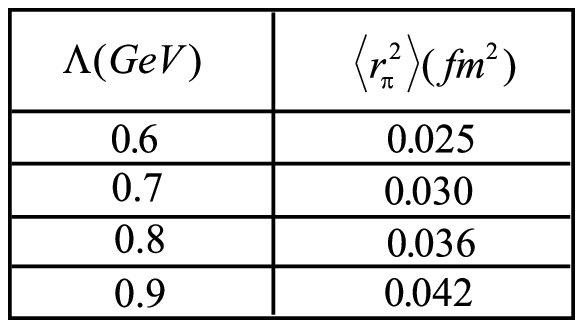,width=7cm}}
\end{center}
\end{figure}
\vspace{-0.7cm}

\begin{center}
{\bf Table II:} \small{The mean square radius $<r_{\pi}^2>$
 for different cut-off values.}
\end{center}

We can see that, whatever the cut-off choice is, the mean 
square radius is much smaller than the experimental value 
$<r_{\pi}^2>=0.439\pm0.008\,fm^2$ \cite{amendolia}.

In Fig. 4 we plot  the form factor $F^+_{\pi}(Q^2)$ without multiplying 
it by $Q^2$ to show in detail the behavior of our curves at low values 
of $Q^2$.

\begin{figure}[h]
\begin{center}
\centerline{\epsfig{figure=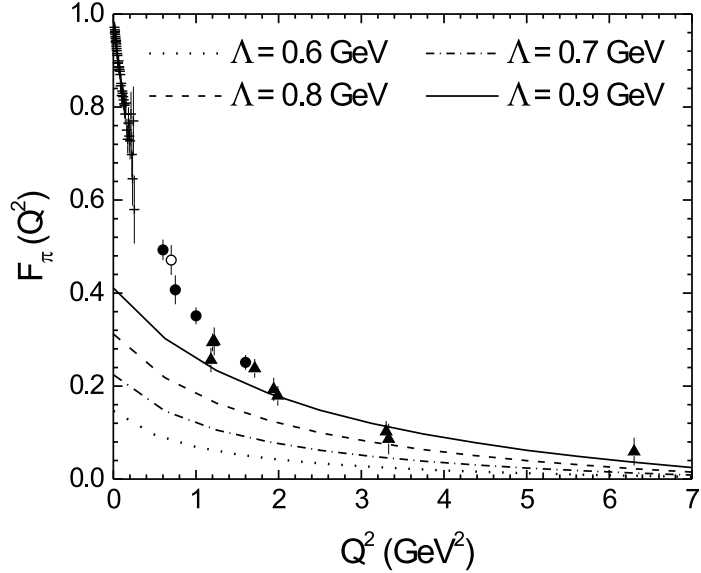,width=11cm}} 
\centerline{\caption{The same as figure 3 for $F_{\pi}(Q^2)$.}}
\end{center}
\label{cutsdifsemxq2}
\end{figure}

The results in Figs. 3 and 4  indicate that the cloud contribution 
underestimates  the data in the low $ Q^2 $ region. At higher values  of $ Q^2 $  
($ Q^2 > 2$ $ GeV^2$),  it starts to be closer to data.   
In the  asymptotic $Q^2$ limit the cloud contribution vanishes.  
Since we did not  
make any model for the bare pion, its influence on the form factor is 
a priori unknown. Looking at the results presented in Fig. 4, we can just observe 
that there is room for a significant bare pion contribution.

Perturbative QCD predicts that in the  high  $ Q^2 $ region  
the form factor should be much smaller than the actually measured data.  
The cloud component is then a candidate to be  {\it the nonperturbative complement to 
pQCD}.  In order to further investigate  the  relation bewteen the cloud 
and  pQCD,  we make now the simplifying assumption that in the asymptotic $Q^2$ region 
the bare pion contribution is simply given by  pQCD. The physical picture becomes quite
simple: a highly virtual photon  probes a pion which is either the "core" or the 
cloud. In the first case the photon interacts with the quarks in the perturbative regime. 
In the second case, the photon interacts with a  hadron of the cloud.
In practice, the above assumption means adding to the cloud a contribution which is known 
and given by \cite{fac}:
\begin{equation}
F_{\pi}(Q^2) =  \frac{16 \pi f^2_{\pi} \alpha_s }{ Q^2}
\label{fqcd}
\end{equation}
where $f_{\pi} = 92.4$ $ MeV$ and $\alpha_s = 0.4$.  This  expression is plotted 
in Fig. 5 with a dotted line and it is added to the previously computed  cloud 
contribution  for $\Lambda = 0.7$ $ GeV$, represented by the dashed line. We  can see 
that the sum of the two contributions (solid line) gives a good description of data 
in the region  $Q^2 > 1$ $ GeV^2$. In Fig. 5 we can clearly observe how, 
at large $ Q^2 $, the cloud contribution dies out while pQCD takes over.  Although our
results are not definitive, they strongly suggest that a hard cut-off 
($ \Lambda > 0.7- 0.8 $ $GeV$) is hard to accomodate, if one is bound to reproduce data 
leaving room for a sizeable bare pion (pQCD) component.
\begin{figure}[h]
\begin{center}
\centerline{\epsfig{figure=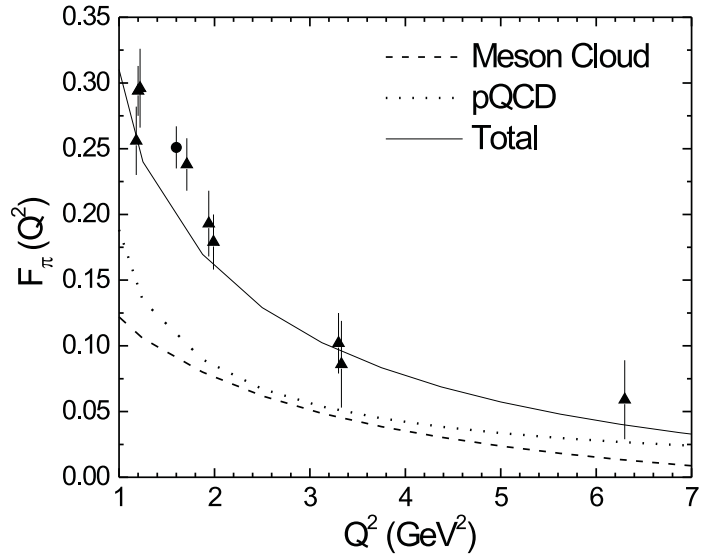,width=11cm}} 
\centerline{\caption{$F_{\pi}(Q^2)$ with cloud ($\Lambda=0.7$ GeV) and pQCD component.}}
\end{center}
\label{pqcdmaisnuvem}
\end{figure}

\section{Concluding remarks}

Meson loop effects are a necessary consequence of quantum
 field theory and must be considered 
in the analysis of any hadronic observable. We have performed
 a straightforward extension of 
standard one loop calculations to the pion form factor.
 The main goal was to estimate the 
magnitude of the meson cloud contribution to $F_{\pi}( Q^2)$.
 It turned out to be less important than
the meson cloud contribution to the nucleon form factor \cite{miller}. 
This is not unexpected since, because of the masses and corresponding virtualities, 
it is much easier for a proton to be in a neutron-pion state than it is for a pion 
to fluctuate into a pion-rho state.

The size of the cloud contribution must also be cross checked with
that in deep inelastic scattering (DIS), as discussed in \cite{speth}.
In Fig. 4 we can see that the size of the 
cloud contribution depends on the value of $Q^2$. At $Q^2$ around zero it 
represents something between $0.15$ and $0.40$ of the total form factor. This fraction is  
consistent with other estimates \cite{speth}, coming from DIS and high energy inclusive 
hadron-hadron collisions. In \cite{speth}, a  smaller percentage was found for the cloud 
contribution. This discrepancy is probably due to the different methods used to extract the 
strength of the cloud states and also because different experimental data were considered.
As $Q^2$ increases, the relative contribution of 
the cloud increases, but this region it is difficult to draw firm conclusions because 
the model gradually runs out of its domain of  
applicability and also, as stated above, our procedure gives an upper 
bound for the cloud contribution.

 The calculation presented here is 
model dependent in the sense that we have truncated a series
 (of cloud states) and replaced the 
effects of the higher order terms by form factors, which also
 account for the quark substructure 
of mesons. Since we have used the same type of form factor and
 the same cut-off for all vertices 
we have only one parameter, $\Lambda$, which was already 
constrained by our previous knowledge 
of  hadronic interactions. It should lie in the interval
 $0.5 < \Lambda < 1.0$ $GeV$. Our 
results confirm this expectation and also suggest that the 
cut-off should rather be in the range  $0.6-0.7$ GeV if we believe that
that the cloud contribution is of the order of $15 \%$  or less. 

The fact that meson loops can not
 describe data in the low $Q^2$ 
region, regardless of cut-off choice, confirms that another 
dynamics is missing in this 
region, as, for example, the interaction of the photon with
 constituent quarks.
\\

\underline{Acknowledgements} This work was partially supported 
by CAPES, CNPq and FAPESP under contract 00/04422-7.

\end{document}